\documentclass[conference]{IEEEtran}
\IEEEoverridecommandlockouts
\usepackage{cite}
\usepackage{amsmath,amssymb,amsfonts}
\usepackage{algorithmic}
\usepackage{graphicx}
\usepackage{textcomp}
\usepackage[table]{xcolor}
\def\BibTeX{{\rm B\kern-.05em{\sc i\kern-.025em b}\kern-.08em
    T\kern-.1667em\lower.7ex\hbox{E}\kern-.125emX}}

\usepackage{multirow}
\usepackage{booktabs}
\usepackage{hyperref}
\usepackage{cleveref}

\crefname{figure}{Fig.}{Figs.}
\crefname{table}{Table}{Tables.}
    
\begin{document}

\title{Chain Reactions in Space: Analyzing the Impact of Satellite Collisions and Debris Accumulation}

\author{\IEEEauthorblockN{Mark Ballard}
\IEEEauthorblockA{The Ohio State University \\
ballard.313@osu.edu}
\and
\IEEEauthorblockN{Guanqun Song}
\IEEEauthorblockA{The Ohio State University \\
song.2107@osu.edu}
\and
\IEEEauthorblockN{Ting Zhu}
\IEEEauthorblockA{The Ohio State University \\
zhu.3445@osu.edu}
}

\maketitle

\begin{abstract}
The exponential increase in artificial satellites, growing from 852 in 2004 to over 9,000 in 2023, has intensified the risk of the Kessler Syndrome: a cascading chain reaction of orbital collisions. This paper analyzes the dynamics of space debris accumulation to identify the primary orbital features contributing to this systemic risk. We compiled and analyzed Two-Line Element (TLE) datasets from Space-Track.org and historical collision data using a Python-based data mining approach. Specifically, we derived satellite velocities using the Vis-Viva equation and evaluated the correlation of five key features, launch piece count, orbital period, apogee, perigee, and Radar Cross Section (RCS) size, with debris density. Our evaluation reveals that \textit{apogee} and \textit{orbital period} exhibit the strongest correlation with the risk of the Kessler Syndrome, indicating that satellites in higher orbits pose a disproportionately greater threat to long-term sustainability due to navigational constraints. Contrary to common assumptions, our data suggests that velocity and object size (RCS) show negligible direct correlation with collision incidence in the current dataset. Based on these findings, we propose mitigation strategies focusing on integrating AI-driven autonomous navigation systems and deploying advanced radiation-resistant shielding materials to enhance the resilience of high-orbit assets.
\end{abstract}

\section{Introduction}
Artificial satellites in space have been growing at an accelerated rate. In 2004, there were only 852 active satellites, a number that skyrocketed to 9,115 as recently as 2023 \cite{statista_satellites}. This represents an average rate of change of 434.89 satellites per year. This exponential growth in satellites is not without consequences. As more satellites enter the atmosphere, the risk of collisions and debris becomes a growing concern. Additionally, the cost of those impacts becomes a concern. Thousands of satellites orbiting in proximity to one another lead to an issue of additional collisions once an impact occurs. This effect is known as the Kessler Effect. 

In 1978, NASA space debris expert Don Kessler observed that once past a certain critical mass, the total amount of space debris will keep on increasing: collisions give rise to more debris and lead to more collisions, in a chain reaction \cite{aiaa_kessler}. After a review of the literature and interviews, a number of experts, ranging from aerospace engineers and planetary scientists to astrodynamicists, show that the scientific community has not reached a consensus about whether Kessler Syndrome has begun, or if it has not begun, how bad it will be when it starts \cite{statista_satellites}. There is an agreement that the domino effect of catastrophic collisions is feasible \cite{space_kessler}. In 2009, a US Iridium Satellite and a defunct Russian military satellite collided and brought more than 2,300 pieces of debris into space \cite{aerospace_debris}. 

The US Iridium Satellite collided with the Russian satellite at a speed of 11.7 km/s (26,000 mph) \cite{aerospace_debris}. The Russian satellite was Kosmos 2251, which was a 950-kilogram (2,100 lb) object, and the Iridium Satellite was a 560-kilogram (1,200 lb) object. The collision took place in Low Earth Orbit (LEO) at 72.52° N, 97.39° E. Fortunately, there were no cascading collisions from this event, and Don Kessler succinctly noted, “while it took 50 years for Earth orbit to become sufficiently congested that we would expect such an event, it will take only about 10 years before another can be expected” \cite{space_crash_blame}. 

It is also important to note that in 2009, there were less than 1,000 satellites (986 total) when this occurred. Now, Starlink reports 1,600 “close encounters” each week, accounting for about 50\% of such incidents \cite{space_starlink_collision}. Starlink is able to record these encounters with its large coverage of LEO satellites. Currently, Starlink is the world’s largest LEO internet constellation \cite{thibault2022leo}. This coverage allows it to deploy a wide array of sensing technology in space. It does this primarily with lasers to communicate between satellites capable of exchanging information within the same area. Starlink has many satellites, wide surface coverage, and extremely low network latency, which is expected to be even lower than traditional optical fiber transmission in the future \cite{thibault2022leo}. While optical links improve speed, they also introduce new complexity in communication reliability \cite{290983}, and optimizing such global constellations is critical for next-generation communication frameworks \cite{gao2024optimizingglobalquantumcommunication}.

Among the approximately 12,000 satellites they have deployed, there are three stages stratified by orbital altitude. These stages help provide coverage from LEO to GEO (Geostationary Earth Orbit), which is the highest orbit where satellites can operate \cite{thibault2022leo}.  The newer satellites are self-monitored, and autonomous systems maneuver the satellites away from possible collisions \cite{techreview_starlink_broken}. To support this autonomy, modern satellites increasingly rely on heterogeneous computing systems \cite{khatri2022heterogeneouscomputingsystems} to process trajectory data onboard. However, the high computational load raises challenges in thermal management, as heat dissipation is difficult in vacuum \cite{yuan2024heatsatellitesmeatgpus}, and requires energy-efficient protocols like LoRaWAN to maintain longevity in LEO environments \cite{shergill2024energyefficientlorawanleo}.

Currently, space traffic coordination systems screen trajectories of spacecraft and objects in space and alert operators on the ground. This occurs when both objects reach a tolerance of close approach beyond the acceptable limit. Coordination is critical between operators to reduce the likelihood of collision. As can be predicted, the process is very convoluted and involves operators sending emails and making calls. A project manager at NASA noted that “Occasionally, we’ll do a maneuver that we find out wasn’t necessary if we could have waited before making a decision. Sometimes you can’t wait three days to reposition and observe. Being able to react within a few hours can make new satellite observations possible.” 

In contrast, Starlink uses onboard communication and navigation to help expedite these maneuvers. This helps with the advent of Starlink set to deploy 12,000 satellites into space. As mentioned previously, Starlink uses satellite-to-satellite connections that are supported with ground station communications. The advent of satellite-to-satellite communication allows them to enable direct communication between satellites, eliminating the need for operators and manual control \cite{techreview_starlink_full}. Ground control is not necessary for maneuvering, but it is integral for monitoring the health and performance of the satellite. The increasing density of satellites and debris in Earth's orbit poses a significant threat to the sustainability of space activities. As the number of satellites continues to grow, so does the risk of collisions. The Starlink satellites use advanced navigation systems, but the problem is these satellites tend to form large “mega constellations” \cite{telecom_starlink}. This can generate additional debris and trigger a cascade of further collisions, known as the Kessler Syndrome. This project is motivated by the urgent need to understand and mitigate these risks to ensure the long-term viability of space operations, aligning with broader goals of achieving carbon neutrality and reducing the environmental impact of device obsolescence \cite{yu2024achievingcarbonneutralityio, gould2024environmentaleconomicimpactio}.

By analyzing the dynamics of satellite collisions and debris interactions, we aim to develop strategies to minimize potential damage and prevent catastrophic chain reactions in space. This research is crucial not only for the protection of valuable space assets but also for the safety and reliability of services that depend on satellite technology, such as communication, navigation, and Earth observation.

\section{Related Work}
\textbf{Kessler Effect:} The danger of a domino effect where more collisions result after two artificial satellites collide is known as the Kessler Effect. To this extent, the debris left in space would cause a chain reaction resulting in endless collisions, causing costly damage and loss of services for many across the globe. To put it more descriptively, Donald Kessler posits that “As the number of artificial satellites in earth orbit increases, the probability of collisions between satellites also increases. Satellite collisions would produce orbiting fragments, each of which would increase the probability of further collisions, leading to the growth of a belt of debris around the earth. This process parallels certain theories concerning the growth of the asteroid belt. The debris flux in such an earth-orbiting belt could exceed the natural meteoroid flux, affecting future spacecraft designs. A mathematical model was used to predict the rate at which such a belt might form. Under certain conditions, the belt could begin to form within this century and could be a significant problem during the next century. The possibility that numerous unobserved fragments already exist from spacecraft explosions would decrease this time interval. However, early implementation of specialized launch constraints and operational procedures could significantly delay the formation of the belt” \cite{kessler1978collision}.

\textbf{Space Debris and Detection:} In recent years there have been many efforts to monitor space debris and have satellites maneuver away from potential danger. Although space is vast, it is also complex. Space debris—commonly known as space junk—refers to non-functional, human-made objects that orbit the Earth and are no longer operational \cite{bonview_ai_space}. This classification includes a diverse array of items, such as discarded rocket stages, inactive satellites, and fragments resulting from collisions or explosions in the space environment \cite{bonview_ai_space}. The management of such obsolete devices has significant environmental and economic impacts \cite{cheng2024technologicalprogressobsolescenceanalyzing}.

There are three different orbital zones. Low Earth Orbit (LEO) is the orbital range closest to Earth and encompasses satellites that orbit up to 1,931 km above Earth \cite{aerospace_orbit101}. It is also the most populated zone. The other two are Medium Earth Orbit (MEO), occupying 1,931 km to 35,888 km, and Geostationary Earth Orbit (GEO), occupying altitudes above MEO \cite{aerospace_orbit101}. About 10\% of all recorded fragmentation debris in the DISCOS database is due to collisions \cite{facchinetti2020governing}. Another 25\% of debris originates from deliberate anti-satellite (ASAT) action, and the remaining 65\% is due to explosions and other non-collision fragmentation events \cite{facchinetti2020governing}. Historically, a variety of techniques have been employed to monitor and identify space debris. These approaches include sophisticated algorithms, such as Particle Swarm Optimization (PSO) and Machine Learning (ML) techniques that aim to estimate debris, assess impact risks, and facilitate mission planning. Most notably, monitoring space debris has become a key area for AI \cite{facchinetti2020governing}. Efficiently processing the vast amount of tracking data often requires advanced techniques like map-reduce for multiprocessing \cite{qiu2023mapreducemultiprocessinglargedata} and precise data classification methods \cite{dixit2023dataclassificationmultiprocessing}, while edge-based semantic segmentation \cite{safavi2023efficientsemanticsegmentationedge} can help satellites identify debris autonomously.

\begin{figure*}[t]
    \centering
    \includegraphics[width=1.0\textwidth]{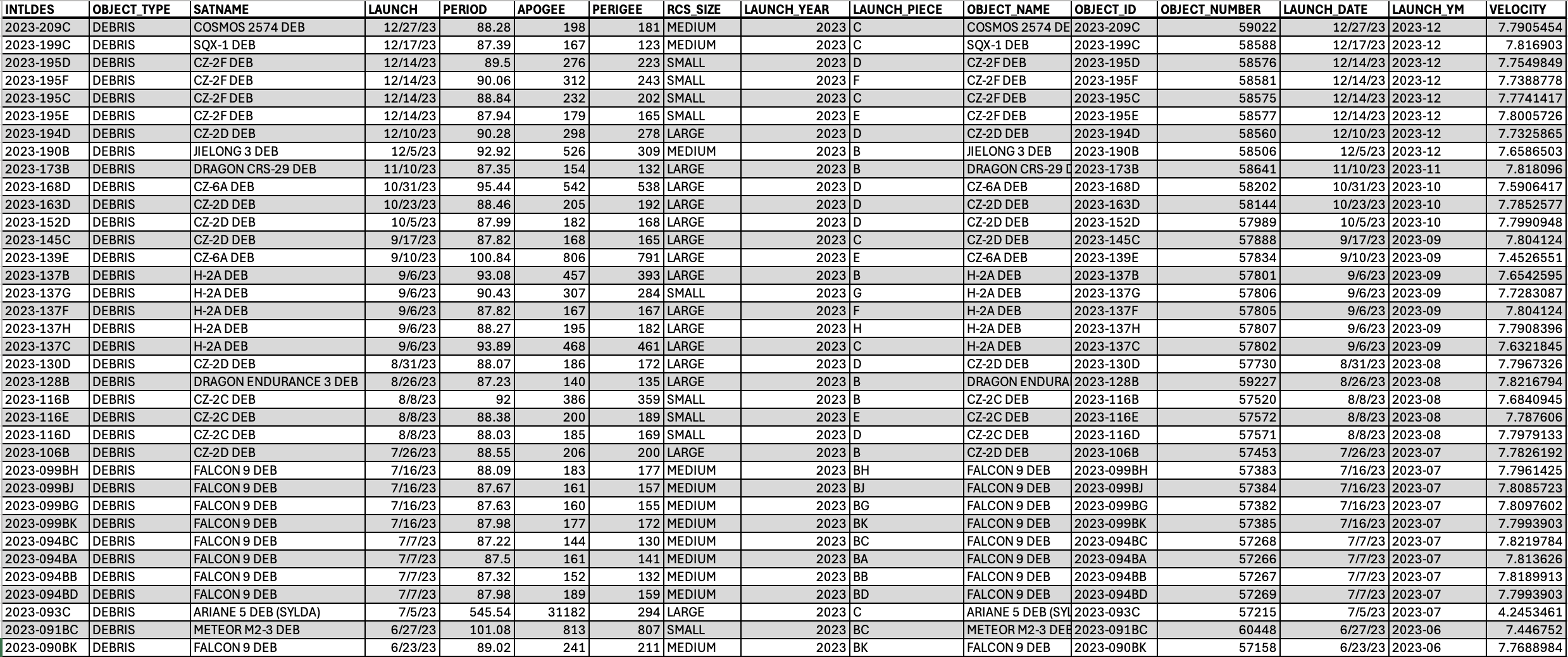}
    \caption{
      TLE Dataset with 5  features showing period, apogee, perigee, RCS size, and velocity
    }
    \label{fig:0}
   
\end{figure*}

\begin{figure}[t]
  \includegraphics[width=1.0\columnwidth]{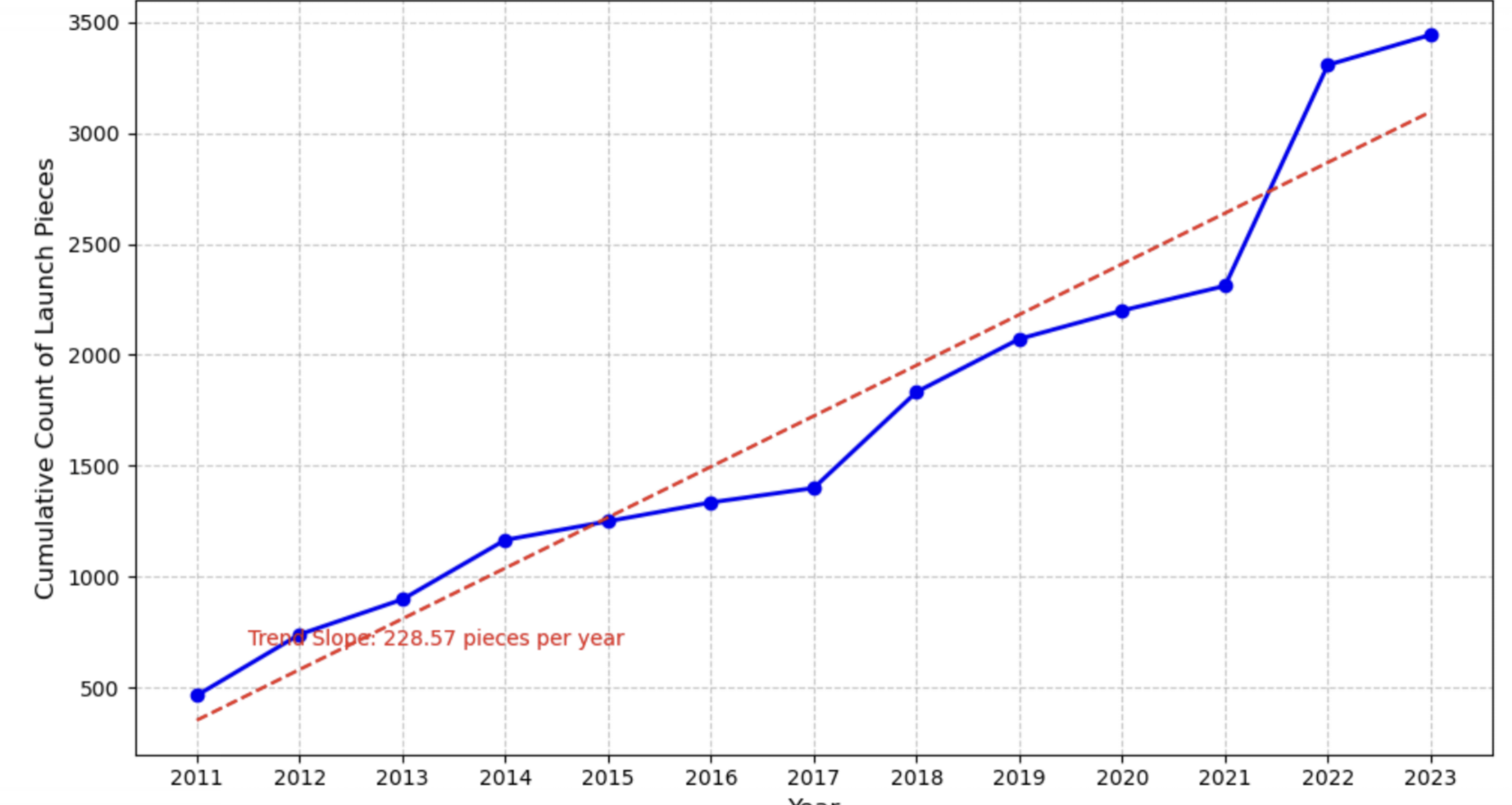}
  \caption{Total launch pieces lost in space over time}
  \label{fig:1}
\end{figure}

\section{Design}
We accessed and compiled comprehensive datasets on current satellite and debris positions, trajectories, and densities from space agencies and organizations dedicated to tracking space objects. Most notably, we used the Two-Line Element (TLE) dataset from Space-Track.org to download the data.  We accessed the data by writing a script in Jupyter Notebooks (Python) to connect and access the publicly available data. TLE is a standardized format used to describe the orbit of a satellite \cite{space_track_tle}. It consists of a set of data that includes important orbital parameters, such as the satellite’s position, velocity, and other relevant information at a specific epoch or time \cite{space_track_tle}. We will also track the mass of satellites in the UCS Satellite Database maintained by the Union of Concerned Scientists. 

For the timeline of space collisions and space junk, we will rely on the Space Debris Environment Report issued by the European Space Agency (ESA) Space Debris Office to get the total number of collisions and their parameters over a span of 12 years from 2011 (when the number of satellites in orbit reached over a thousand) to 2023 (the most recent data reported for the number of satellites currently in space). We will collect the data and perform a risk analysis in Jupyter Notebooks using Python. We will also create a diagram in Jupyter Notebooks to map out the current risk of collision in space based on orbital patterns. We will also develop a heatmap of the latest known features for the entire dataset. Designing a histogram to forecast events from the database presents a few challenges. First, the distribution of the final risk associated with all the conjunction events contained in the database could be highly skewed \cite{bonview_ai_space}. Secondly, there is significant heterogeneity in the various time series associated with different events \cite{bonview_ai_space}. Lastly, there will be an assumption of satellites having single encounters with objects in space, but it is possible that multiple encounters could happen with the same object \cite{bonview_ai_space}. It is important to note, though, that those encounters can be assumed to be rare. All in all, the heatmap should have some utility in giving a general idea of whether or not the Kessler effect is fast becoming a reality. This will help us to extrapolate the data to predict the future consequences of accumulating space debris. From this data, we will formulate recommendations, policy changes, or regulatory actions to mitigate risks for future space operations.

\section{Evaluation}
We approached interpreting the number of space collisions per year in three steps. The first step is to obtain publicly available data in order to represent the number of space collisions a year. Secondly, since actual space events are rare, we interpreted space collisions to include satellite breakup events or fragmentation events, which we classified appropriately as space debris. Third, we formulate recommendations, policy changes, or regulatory actions to mitigate risks for future space operations. 
Our implementation of using the dataset of the space debris uses Jupyter Notebooks to use linear regression methods to uncover new insights. Our implementation runs entirely in Jupyter Notebooks using Python. We will use data mining methods such as correlation plots, line plots, and data imported from free publicly available data from U.S. Space-Track system reports. We will also collect information from sites that are in the public domain.

\subsection{Analysis of the Space Debris Dataset}
We first created an account on Space-Track to obtain the ability to access the database. Next, we reached out to researchers that were analyzing the data and obtained sample TLE data. The TLE data showed the satellite names and debris. However, we needed the data to be more granular, so we modified the type of debris that was created and the size of the debris. We were interested in obtaining the velocity for the satellites; however, that was not publicly available. It is important to note that in space, all satellites orbit at around the same velocity with little deviation. Our research revealed that the average speed of satellites in space is approximately 7.8 km/s \cite{space_leo}. To verify this, we use a simplified version of the Kepler equation to derive the velocity for each satellite. The complete version of the equation can be found in Kessler’s first paper on the danger of the domino effect of satellite collisions. We found a shorter version called the Vis-Viva Equation (1), from Kepler’s second law, to help us derive the speed of each satellite \cite{fiu_orbits}:

\[
v^2 = \mu\left(\frac{2}{r} - \frac{1}{\alpha}\right)
\]

\begin{figure}[t]
  \includegraphics[width=1.0\columnwidth]{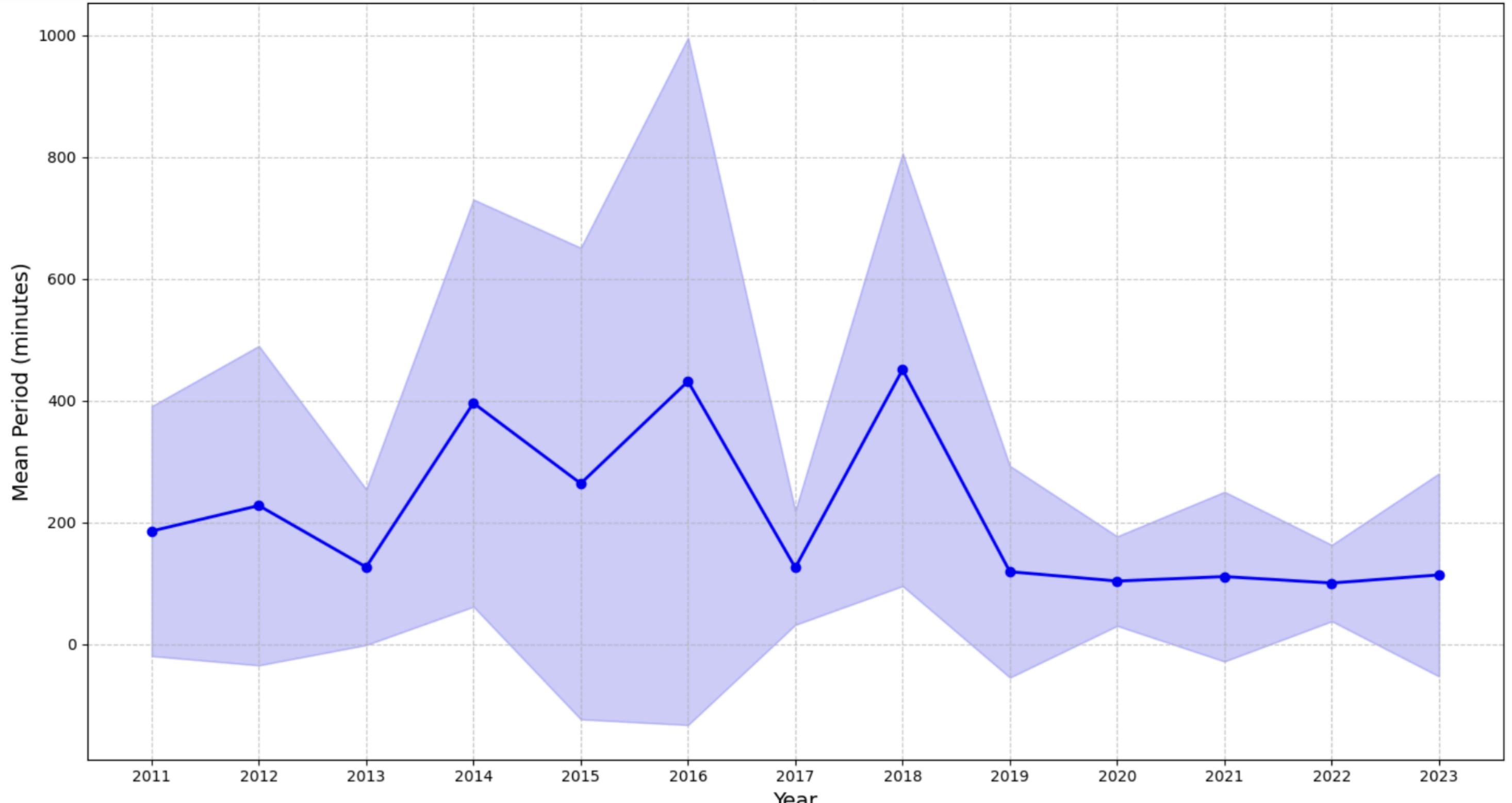}
  \caption{Mean, minimum, and maximum orbital period each}
  \label{fig:2}
\end{figure}

Further deriving the equation by taking the square root will give us the velocity, $v$. We will now explain the other variables: $\mu$ is the Earth’s gravitational parameter \cite{fiu_orbits}; $r$ is the magnitude of the position vector and is the distance between the two bodies and the orbit’s semi-major axis \cite{fiu_orbits}; $\alpha$ is the semi-major axis and is half the length of the largest diameter of the ellipse, called the major axis. After deriving this formula, we used the measurements in the dataset to calculate our velocity using Python. We wrote a script in Python to modify the TLE dataset to add these features shown in Fig. \ref{fig:0}.

\begin{figure}[t]
  \includegraphics[width=1.0\columnwidth]{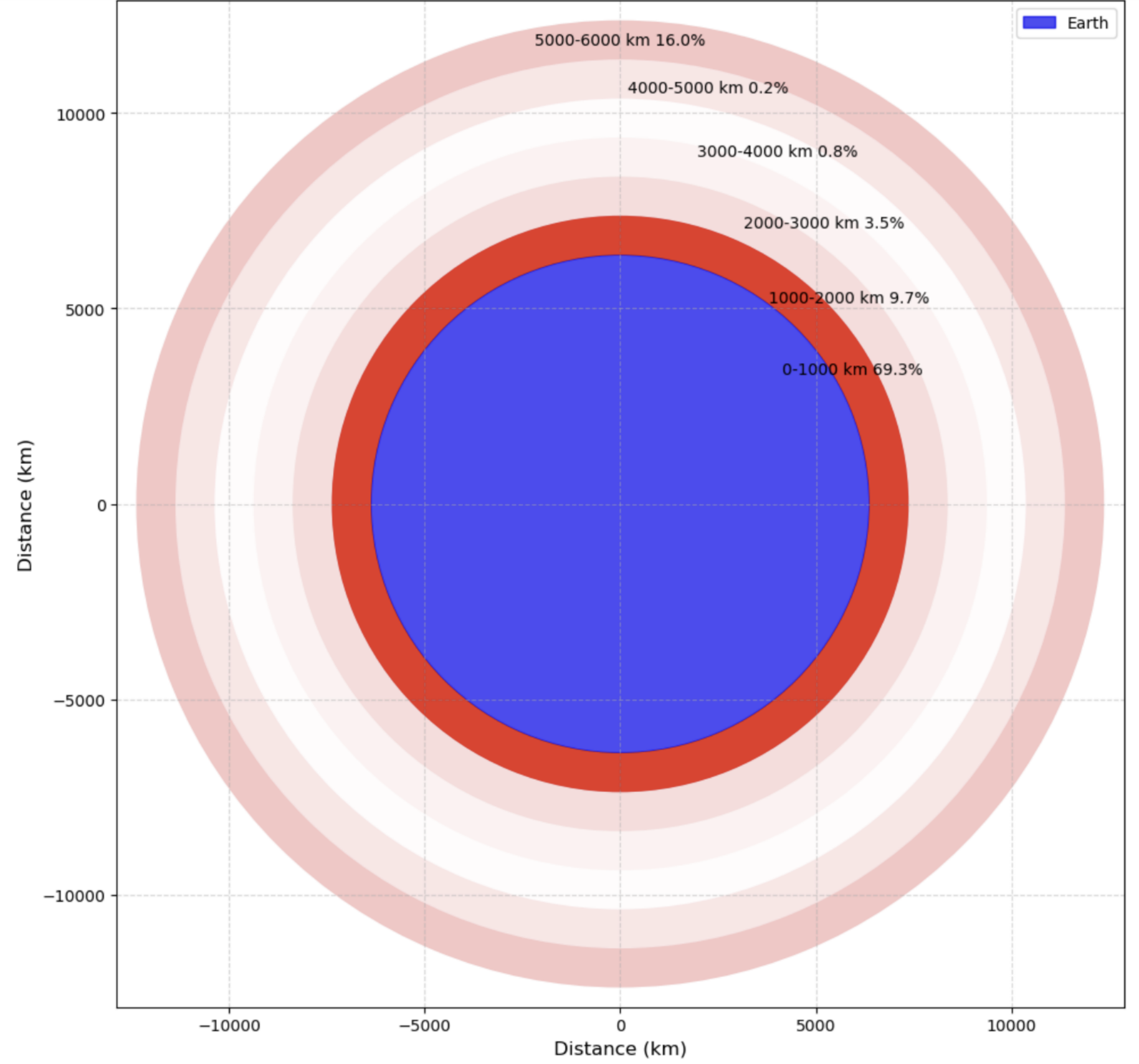}
  \caption{Graph in polar representation of space debris density at each altitude}
  \label{fig:3}
\end{figure}

\begin{figure}[t]
  \includegraphics[width=1\columnwidth]{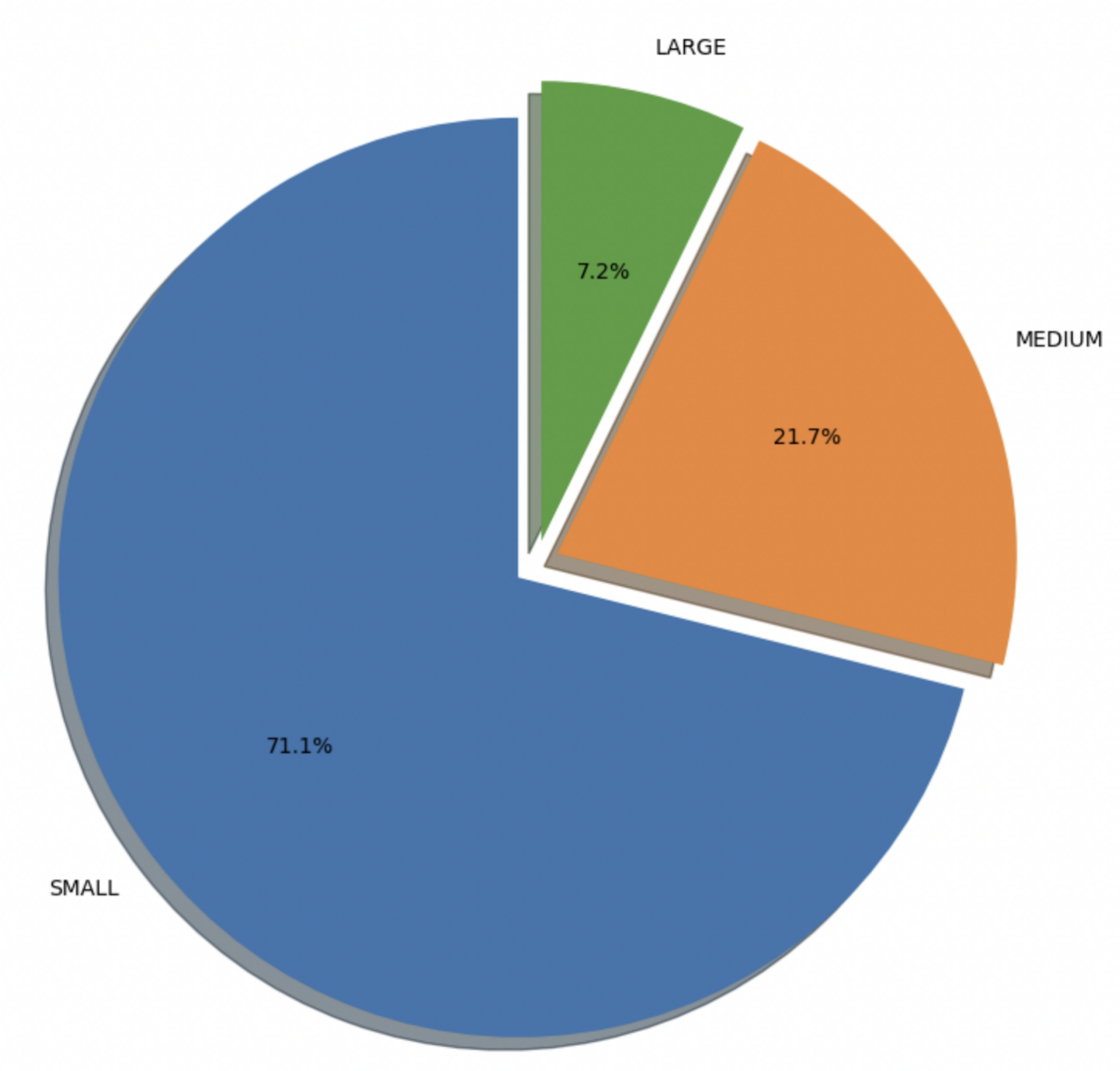}
  \caption{Distribution of RCS Size of Space Debris}
  \label{fig:4}
\end{figure}

\begin{figure}[t]
  \includegraphics[width=1.0\columnwidth]{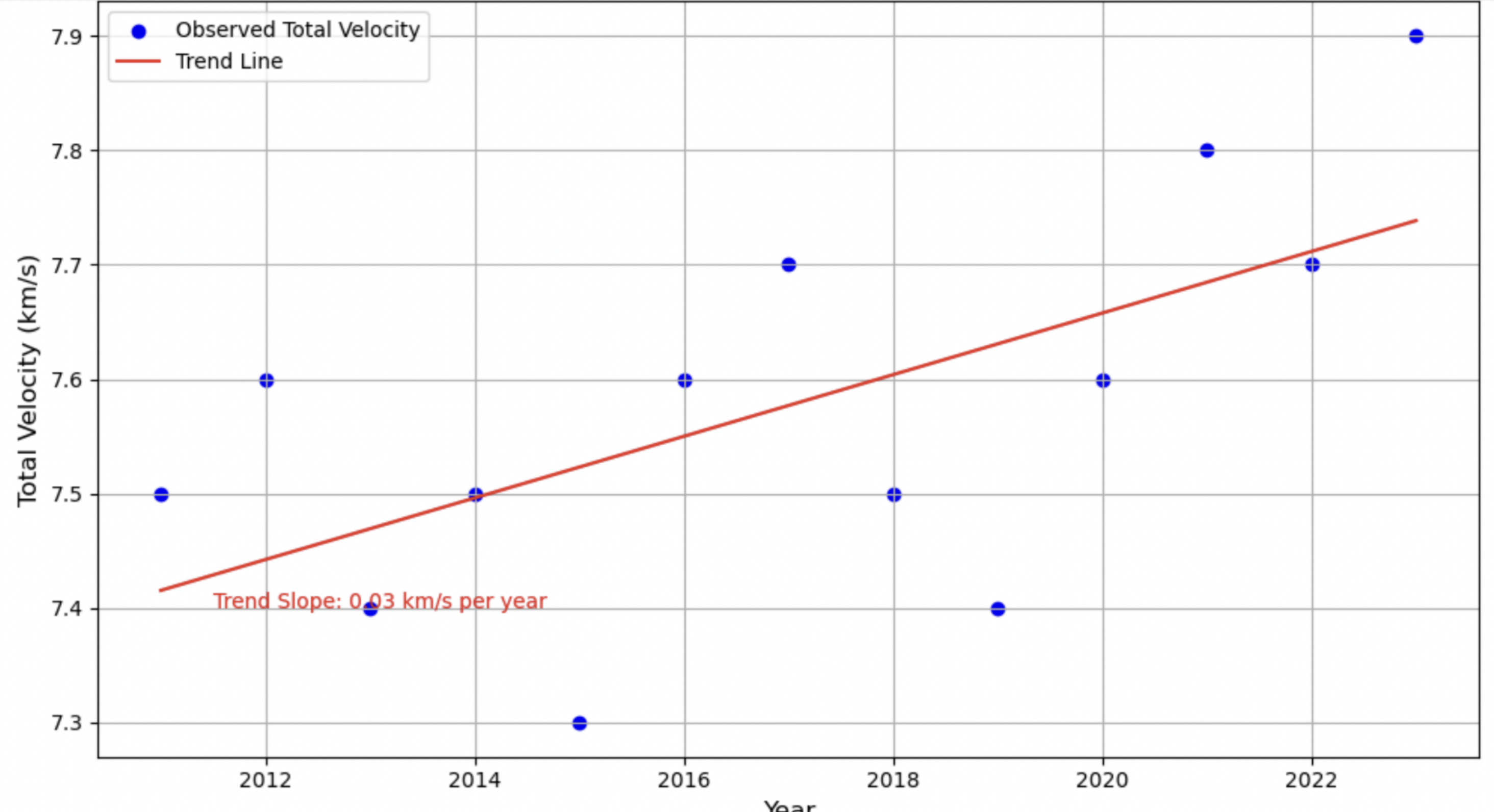}
  \caption{Trendline of  total velocity of space debris over time}
  \label{fig:5}
\end{figure}

Using these five key features—launch piece, period, apogee, perigee, and Radar Cross Section (RCS) size—from the debris count database, we analyzed the data to see which of these features had the highest relation to large debris count. Specifically, launch piece refers to the fragment expelled from the satellite while entering orbit. Fig. \ref{fig:1} shows the cumulative total of launch pieces in space. The period is the orbital period or the time it takes for the debris to complete one orbit around Earth. Fig. \ref{fig:2} shows the mean orbital period by year and the fluctuation from the mean. The dataset ranges from 86 minutes as the minimum and 11,000 minutes (8 days) as the max. Apogee and perigee refer to the farthest and closest distance of that object to Earth, respectively. Fig. \ref{fig:3} is a visualization of the density of debris from collisions in space. The visualization shows that most debris are within 0-1000 km (69.3\%) from Earth’s surface (blue circle). A moderate proportion of the debris is found at 1000-2000 km (9.7\%). The proportion got progressively smaller at 2000-5000 km, and 5000–6000 km had the second largest proportion (16\%). This may be because most satellites commonly operate at the LEO and GEO layers \cite{interseas_leo_geo}. RCS size is the size of that object in space. Fig. \ref{fig:4} shows the proportion of each classification of RCS-sized objects in space: small (smaller than 0.1 $m^2$), medium (0.1 $m^2$ – 1 $m^2$), and large (greater than 1 $m^2$) \cite{space_track_rcs}. We also used the velocity of each debris piece as a feature. Our reasoning is that velocity is a good predictor of the frequency and impact of collisions, so we used that as our focus \cite{liou2020impact}. Fig. \ref{fig:5} clearly shows that there has been a trend of velocity of space debris increasing over time.

\begin{figure}[t]
  \includegraphics[width=1.0\columnwidth]{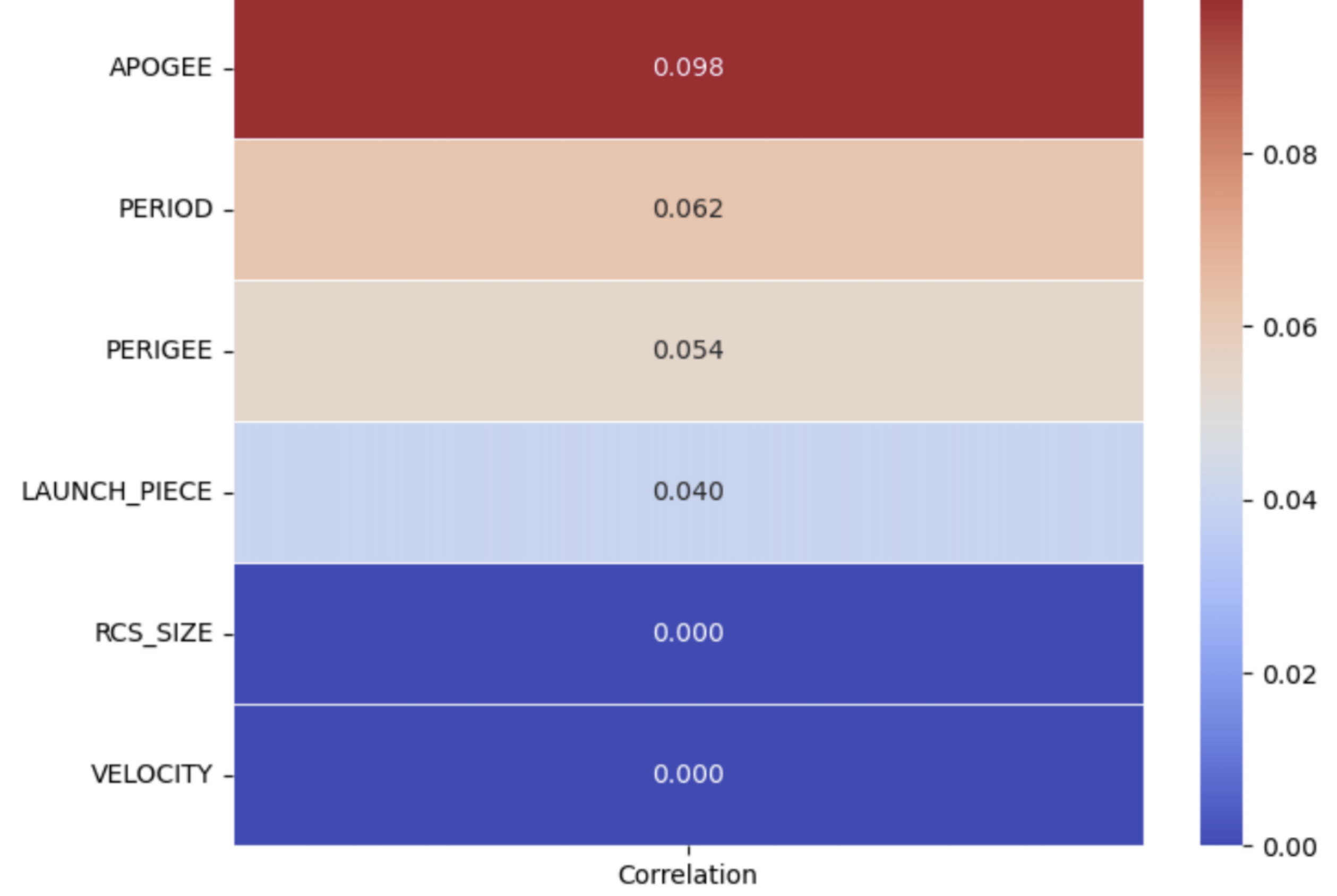}
  \caption{Heatmap of all features to correlation with space collision}
  \label{fig:7}
\end{figure}

\subsection{Features Comparison}

Another analysis of the five features to the odds of the Kessler syndrome was conducted in Fig. \ref{fig:7}. This shows that of the five features, apogee had the highest correlation to the risk of the Kessler syndrome. We also included the total velocity to see if it had an impact. We discovered that apogee and then the orbital period had more of an impact on the risk of the Kessler syndrome than even total velocity. This shows that satellites at high orbits pose more of a risk in increasing the Kessler Syndrome. Period also correlates with the apogee because the higher the satellite is in orbit, the longer its orbital period. The findings are summarized below in Fig. \ref{fig:7}. The correlation for each feature was found by finding the Pearson correlation coefficient for each feature and total collisions. It is found by dividing the covariance of each feature to collisions and the product of their standard deviation. As far as the data suggests, it can be reasonably assumed that velocity and RCS size showed no correlation with space collision; the incidence of a collision showed no noticeable increase or decrease of risk for any change in velocity or RCS size.

\section{ISSUES WE ENCOUNTERED}
The issues we encountered on this project were the nature of the Kessler Syndrome not having a universally agreed starting point \cite{statista_satellites}. We had to use our understanding of the features of collisions as our focus on determining what should be the focus of our recommendations. Also, some of the data for the satellites was not publicly available, so we had to derive it. Namely, the velocity of the satellites in orbit was calculated using Kepler’s second law. Operators and automated systems would monitor the status of the satellite and would have that data. However, we feel this would not affect our recommendations.

\section{CONCLUSION}
Collisions are complex events and hard to observe in the conventional sense. Satellite collisions show an increasing trend over the years, and this poses a risk of the Kessler Syndrome. Although there is no consensus on the Kessler Syndrome and what could cause it, we can use six key features to help give us a sense of the risk. As far as the data suggests, satellites with higher orbits and long orbital periods pose a risk for the Kessler Syndrome. This may be due to the fact that satellites at higher orbits are more difficult to control and navigate \cite{interseas_leo_geo}. Long periods in space may contribute to the risk of satellites malfunctioning and breaking apart. Our recommendations would be to increase the navigational ability of satellites at higher orbits, which would require more sensitive sensors and more intelligent systems on board. The advent of smarter systems such as AI could ameliorate this. We also recommend a more robust design of satellites so that the sensitive systems are shielded from the harsh environments in space. Advances in radiation-resistant materials, thermal management solutions, impact-resistant materials, and self-healing technologies would help to create more resilient and long-lasting satellites. These recommendations could not only protect valuable space assets from collisions but also improve the safety and reliability of services for the future of space technology.

\bibliography{reference,zhu}

@misc{statista_satellites,
  author = {{Statista Research Department}},
  title = {Number of active satellites from 1957 to 2022},
  year = {2023},
  url = {https://www.statista.com/statistics/897719/number-of-active-satellites-by-year/},
  note = {Accessed: 2024}
}

@misc{aiaa_kessler,
  title = {Understanding the Misunderstood Kessler Syndrome},
  author = {{Aerospace America}},
  url = {https://aerospaceamerica.aiaa.org/features/understanding-the-misunderstood-kessler-syndrome/},
  year = {2024},
  note = {Accessed: 2024}
}

@misc{space_kessler,
  title = {Kessler Syndrome and the space debris problem},
  author = {Mike Wall},
  journal = {Space.com},
  year = {2021},
  url = {https://www.space.com/kessler-syndrome-space-debris},
  note = {Accessed: 2024}
}

@misc{aerospace_debris,
  title = {A Brief History of Space Debris},
  author = {{The Aerospace Corporation}},
  url = {https://aerospace.org/article/brief-history-space-debris},
  year = {2023}
}

@misc{space_crash_blame,
  title = {Satellite Crash: Who's to Blame?},
  author = {Leonard David},
  journal = {Space.com},
  year = {2009},
  url = {https://www.space.com/4312-satellite-crash-blame.html}
}

@misc{space_starlink_collision,
  title = {SpaceX Starlink satellite collision alerts on the rise},
  author = {Tereza Pultarova},
  journal = {Space.com},
  year = {2023},
  url = {https://www.space.com/spacex-starlink-satellite-collision-alerts-on-the-rise}
}

@article{thibault2022leo,
  title = {LEO Mega Constellations: Review of Development, Impact, Surveillance, and Governance},
  author = {Thibault, C. and others},
  journal = {ResearchGate Preprint},
  year = {2022},
  url = {https://www.researchgate.net/publication/362378005}
}

@misc{techreview_starlink_broken,
  title = {One of SpaceX’s Starlink satellites almost collided with a weather satellite (Archive)},
  author = {Neel V. Patel},
  journal = {MIT Technology Review},
  year = {2019},
  url = {https://www.technologyreview.com/2019/09/02/133180/one-of-spacexs-starlink-satellites-}
}

@misc{techreview_starlink_full,
  title = {One of SpaceX’s Starlink satellites almost collided with a weather forecasting satellite},
  author = {Neel V. Patel},
  journal = {MIT Technology Review},
  year = {2019},
  url = {https://www.technologyreview.com/2019/09/02/133180/one-of-spacexs-starlink-satellites-almost-collided-with-a-weather-forecasting-satellite/}
}

@misc{telecom_starlink,
  title = {The Technology Behind Starlink Satellites},
  author = {{Telecom World}},
  url = {https://telecomworld101.com/the-technology-behind-starlink-satellites/},
  year = {2023}
}

@article{kessler1978collision,
  title = {Collision frequency of artificial satellites: The creation of a debris belt},
  author = {Kessler, Donald J and Cour-Palais, Burton G},
  journal = {Journal of Geophysical Research: Space Physics},
  volume = {83},
  number = {A6},
  pages = {2637--2646},
  year = {1978},
  publisher = {Wiley Online Library},
  doi = {10.1029/JA083iA06p02637}
}

@article{bonview_ai_space,
  title = {Artificial Intelligence for Space Debris Avoidance},
  journal = {Artificial Intelligence and Applications},
  year = {2023},
  url = {https://ojs.bonviewpress.com/index.php/AIA/article/view/3741/1207}
}

@misc{aerospace_orbit101,
  title = {Earth Orbit 101},
  author = {{Aerospace Security Project}},
  url = {https://aerospace.csis.org/aerospace101/earth-orbit-101/},
  year = {2023}
}

@article{facchinetti2020governing,
  title = {Governing the Space Commons},
  author = {Facchinetti, C.},
  journal = {International Journal of the Commons},
  year = {2020},
  url = {https://thecommonsjournal.org/articles/10.5334/ijc.1275}
}

@misc{space_track_tle,
  title = {Space-Track Documentation: TLE},
  author = {{Space-Track.org}},
  url = {https://www.space-track.org/documentation#/tle},
  year = {2024}
}

@misc{space_leo,
  title = {Low Earth Orbit: Definition, Elliptical Orbits and More},
  author = {Elizabeth Howell},
  url = {https://www.space.com/low-earth-orbit},
  year = {2023}
}

@misc{fiu_orbits,
  title = {Orbits (Course Material)},
  author = {{Florida International University}},
  url = {https://faculty.fiu.edu/~vanhamme/ast3213/orbits.pdf},
  note = {Accessed: 2024}
}

@misc{space_track_rcs,
  title = {Space-Track Documentation: Radar Cross Section (RCS)},
  author = {{Space-Track.org}},
  url = {https://www.space-track.org/documentation/loadLegendRCS},
  year = {2024}
}

@article{liou2020impact,
  title = {The impact of large satellite constellations on space debris environment},
  author = {Liou, J.-C. and others},
  journal = {Acta Astronautica},
  year = {2020},
  publisher = {Elsevier},
  url = {https://www.sciencedirect.com/science/article/pii/S0094576520301235}
}

@misc{interseas_leo_geo,
  title = {LEO and GEO Satellites: Differences, Advantages and Challenges},
  author = {{Interseas}},
  url = {https://interseas.es/en/leo-and-geo-satellites-differences-advantages-and-challenges-in-satellite-connectivity/},
  year = {2023}
}

@inproceedings {290983,
author = {Xin Liu and Wei Wang and Guanqun Song and Ting Zhu},
title = {{LightThief}: Your Optical Communication Information is Stolen behind the Wall},
booktitle = {32nd USENIX Security Symposium (USENIX Security 23)},
year = {2023},
isbn = {978-1-939133-37-3},
address = {Anaheim, CA},
pages = {5325--5339},
url = {https://www.usenix.org/conference/usenixsecurity23/presentation/liu-xin},
publisher = {USENIX Association},
month = aug
}

@misc{safavi2023efficientsemanticsegmentationedge,
      title={Efficient Semantic Segmentation on Edge Devices}, 
      author={Farshad Safavi and Irfan Ali and Venkatesh Dasari and Guanqun Song and Ting Zhu and Maryam Rahnemoonfar},
      year={2023},
      eprint={2212.13691},
      archivePrefix={arXiv},
      primaryClass={cs.CV},
      url={https://arxiv.org/abs/2212.13691}, 
}

@misc{khatri2022heterogeneouscomputingsystems,
      title={Heterogeneous Computing Systems}, 
      author={Dimple P. Khatri and Guanqun Song and Ting Zhu},
      year={2022},
      eprint={2212.14418},
      archivePrefix={arXiv},
      primaryClass={eess.SY},
      url={https://arxiv.org/abs/2212.14418}, 
}

@misc{shergill2024energyefficientlorawanleo,
      title={Energy Efficient LoRaWAN in LEO Satellites}, 
      author={Muskan Shergill and Zach Thompson and Guanqun Song and Ting Zhu},
      year={2024},
      eprint={2412.20660},
      archivePrefix={arXiv},
      primaryClass={cs.ET},
      url={https://arxiv.org/abs/2412.20660}, 
}

@misc{gould2024environmentaleconomicimpactio,
      title={Environmental and Economic Impact of I/O Device Obsolescence}, 
      author={Patrick Gould and Guanqun Song and Ting Zhu},
      year={2024},
      eprint={2412.20655},
      archivePrefix={arXiv},
      primaryClass={cs.CY},
      url={https://arxiv.org/abs/2412.20655}, 
}

@misc{gao2024optimizingglobalquantumcommunication,
      title={Optimizing Global Quantum Communication via Satellite Constellations}, 
      author={Yichen Gao and Guanqun Song and Ting Zhu},
      year={2024},
      eprint={2501.00280},
      archivePrefix={arXiv},
      primaryClass={quant-ph},
      url={https://arxiv.org/abs/2501.00280}, 
}

@misc{yuan2024heatsatellitesmeatgpus,
      title={Heat: Satellite's meat is GPU's poison}, 
      author={Zhehu Yuan and Jinyang Liu and Guanqun Song and Ting Zhu},
      year={2024},
      eprint={2501.14757},
      archivePrefix={arXiv},
      primaryClass={cs.DC},
      url={https://arxiv.org/abs/2501.14757}, 
}

@misc{dixit2023dataclassificationmultiprocessing,
      title={Data Classification With Multiprocessing}, 
      author={Anuja Dixit and Shreya Byreddy and Guanqun Song and Ting Zhu},
      year={2023},
      eprint={2312.15152},
      archivePrefix={arXiv},
      primaryClass={cs.LG},
      url={https://arxiv.org/abs/2312.15152}, 
}

@misc{yu2024achievingcarbonneutralityio,
      title={Achieving Carbon Neutrality for I/O Devices}, 
      author={Botao Yu and Guanqun Song and Ting Zhu},
      year={2024},
      eprint={2501.14774},
      archivePrefix={arXiv},
      primaryClass={cs.CY},
      url={https://arxiv.org/abs/2501.14774}, 
}

@misc{cheng2024technologicalprogressobsolescenceanalyzing,
      title={Technological Progress and Obsolescence: Analyzing the Environmental Economic Impacts of MacBook Pro I/O Devices}, 
      author={Yun-Chieh Cheng and Yu-Tong Shen and Guanqun Song and Ting Zhu},
      year={2024},
      eprint={2501.14758},
      archivePrefix={arXiv},
      primaryClass={cs.CY},
      url={https://arxiv.org/abs/2501.14758}, 
}

@misc{qiu2023mapreducemultiprocessinglargedata,
      title={Map-Reduce for Multiprocessing Large Data and Multi-threading for Data Scraping}, 
      author={Zefeng Qiu and Prashanth Umapathy and Qingquan Zhang and Guanqun Song and Ting Zhu},
      year={2023},
      eprint={2312.15158},
      archivePrefix={arXiv},
      primaryClass={math.NA},
      url={https://arxiv.org/abs/2312.15158}, 
}
\bibliographystyle{ieeetr}

\end{document}